\documentclass{WileyMSP-template}
\usepackage{times}
\usepackage{xcolor}
\usepackage[colorlinks,allcolors=blue]{hyperref}
\usepackage{bm}
\usepackage{ulem}
\usepackage{dsfont}
\usepackage{amsmath}
\usepackage{amsfonts}
\usepackage{amssymb}
\usepackage{braket}
\usepackage{textcomp}
\usepackage[utf8]{inputenc}

\definecolor{ao(english)}{rgb}{0.0, 0.5, 0.0}

\newcommand{\muB}{{\ensuremath{\mu_{\mathrm{B}}}}}

\begin{document}

\title{Path integral spin dynamics for quantum paramagnets}

\maketitle

\author{Thomas Nussle$^{(a)}$}
\author{Pascal Thibaudeau$^{(b)}$}
\author{Stam Nicolis$^{(c)}$}

\begin{affiliations}
    $^{(a)}$
    School of Physics and Astronomy, University of Leeds, United Kingdom\\
    t.s.nussle@leeds.ac.uk

    $^{(b)}$
    CEA -- DAM -- Le Ripault, F-37260, Monts, France\\
    pascal.thibaudeau@cea.fr

    $^{(c)}$
    Institut Denis Poisson, Université de Tours, Université d'Orléans, CNRS (UMR7013), Parc de Grandmont, F-37200, Tours, France\\
    stam.nicolis@lmpt.univ-tours.fr

\end{affiliations}


\begin{abstract}
    A path integral method, combined with atomistic spin dynamics simulations, has been developed to calculate thermal quantum expectation values using a classical approach. In this study, we show how to treat Hamiltonians with non-linear terms, that are relevant for describing uniaxial anisotropies and mechanical constraints. These interactions can be expressed solely through quadratic terms of the spin operator along one axis, that can be identified with the quantisation axis.
\end{abstract}

\section{Introduction}
\label{sec:introduction}

Existing numerical methods for describing the properties of magnetic materials, like first-principles approaches and spin models, have limited applicability, when certain constraints must be taken into account. 
In particular, spin models for magnetic materials generally can be treated quantitatively only by imposing quite severe approximations, whose range of validity is not always easy to identify. 
This is particularly the case for understanding how classical behaviour emerges from the microscopic quantum dynamics. 
This is becoming relevant, as magnetic devices become so small that this crossover has practical implications. 
``Small'' here refers to scale, not, necessarily, to the number of degrees of freedom.

On the one hand, quantum models provide precise descriptions of both thermodynamics and dynamics, including quantum effects like entanglement~\cite{laflorencieQuantumEntanglementCondensed2016} and fluctuations~\cite{nelsonQuantumFluctuations1985}. 
However, they are limited to studying ``small'' systems (tens or hundreds of spins) due to high computational costs of current algorithms. 
Quantum Monte Carlo (QMC) can handle large quantum spin systems (hundreds of thousands of spins) accurately but cannot access dynamical quantities since it is a thermodynamic description, which describes equilibrium properties and cannot take into account real time evolution~\cite{foulkesQuantumMonteCarlo2001}. 
Other quantum methods that do describe real-time dynamics, however, cannot handle such ``large'' systems~\cite{andersSpinPrecessionRealtime2006}. 

On the other hand, classical spin models are commonly used to study the dynamics and thermodynamics of magnetic materials, especially at ``higher'' temperatures, where quantum effects, like entanglement can be ignored~\cite{erikssonAtomisticSpinDynamics2017}.
These models have (relatively) low computational cost, are easy to parallelise, and can simulate dynamics for hundreds of thousands or even millions of interacting spins~\cite{tranchidaMassivelyParallelSymplectic2018}.
Although they do provide a good qualitative description of magnetic dynamics, they may struggle at ``lower'' temperatures, when classical Boltzmann statistics is no longer valid and the quantum nature of magnetism becomes relevant.
When the magnon Debye temperature is ``high'', or comparable to the magnetic ordering temperature, the ``low-temperature" quantum regime may in fact encompass most of the magnetic ordering temperature range. 
This means that it becomes inappropriate to use classical statistics in such cases~\cite{barkerSemiquantumThermodynamicsComplex2019}.
The focus for taking into account quantum effects in atomistic spin dynamics is thus centered on discovering analytical expressions and providing a physical explanation for the power–law rescaling that occurs between quantum and classical fluctuations~\cite{berrittaAccountingQuantumEffects2023}.

In this paper, we present a multiscale modelling strategy that relates these two approaches, aiming to fill (some of) the existing gaps. 
Our objective is to set up a framework for studying both thermodynamic equilibrium and nonequilibrium properties of magnetic materials across length scales reaching up to micrometers, starting from fundamental principles.
To effectively design and manage the magnetism of such quantum devices, it becomes essential to understand how the junction's structural and chemical environment impacts the spin center.
To this end, a comprehensive and quantum thermodynamically consistent theory has been developed to describe the quantum dynamics of spins in magnetic materials, including non-Markovian damping, coloured noise and quantum zero-point fluctuations, in a manner analogous to macroscopic quantum electrodynamics, which successfully describes quantum electromagnetism in dielectric materials, without necessarily requiring more detailed microscopic models.
\cite{andersQuantumBrownianMotion2022}. 
Unfortunately, this theory was exemplified only on a single quantum spin in interaction with various bosonic baths.
When the focus is on molecular magnets and single spin centers that can be individually controlled when connected to contacts that form an electrical junction, previous efforts were devoted to the simplest nontrivial spin center: a single spin under an external magnetic field, described by the Zeeman Hamiltonian~\cite{nussleNumericalSimulationsSpin2023}. 
The objective of this study is to investigate the thermodynamic properties of these centers when exposed to more complex external environments.

\section{The paramagnetic Hamiltonian}
\label{sec:definition}
We consider a single spin, described by a (vector) quantum operator $\hat{\bm S},$ that is subject to linear and quadratic local interactions. 
The energy of such a spin is described by a quantum Hamiltonian operator $\hat{\cal{H}}$ that contains a Zeeman interaction with a constant magnetic field ${\bm H}$, a uniaxial anisotropy interaction with an easy axis ${\bm n}$ and intensity $K$, and a magneto-elastic interaction with a stress tensor $\sigma$ of rank 2, a coupling tensor $B$ and mechanical constants tensor $C$, each of rank 4 \cite{dutremoletdelacheisserieMagnetostrictionTheoryApplications1993}.
The symmetries that dictate the form of the Hamiltonian are rotation invariance, which implies that the most general expression is given by the form
\begin{align}
    \label{fullH}
    \hat{\cal{H}}&=-g\muB\mu_0{\bm H}\cdot\hat{\bm S}-K\left({\bm n}\cdot{\hat{\bm S}}\right)^2-B_{IJKL}C^{-1}_{IJMN}\sigma_{MN}\hat{S}_K\hat{S}_L,  
\end{align}
where $g$ is the Landé $g$-factor for the spin and $\muB$ the Bohr's magneton.
Computing the thermal quantum expectation values of functions, $f(\hat{\bm S}),$ of the spin operator(s),
\begin{align}
\label{vevfofS}
\langle f(\hat{\bm S})\rangle\equiv
\frac{\mathrm{Tr}\,f(\hat{\bm S})\,e^{-\beta\hat{\cal{H}}}}
     { \mathrm{Tr}\,e^{-\beta\hat{\cal{H}}}}
\end{align}
in closed form, in the general case, is (a) tedious and (b) not particularly illuminating. The fundamental reason is that the spin operators do {not} commute, since they define a curved target space. This noncommutativity is purely geometric and does not have anything to do with quantum effects. 

However there exists a subclass of interactions, where this noncommutativity is absent and the calculation becomes tractable: This occurs when only one component of the spin operator is present and its direction can be identified with the quantisation $z$-axis; the simplest example, when only the Zeeman interaction was present, was studied in ref.~\cite{nussleNumericalSimulationsSpin2023}. 

Keeping the all the terms of eq.~(\ref{fullH}) in this approximation, one finds 
\begin{align}
    \begin{aligned}
        \hat{\cal H} & =-{g\muB\mu_0 H}\hat{S_z} - K\hat{S}_z^2 -\lambda(\mathds{1}-\hat{S}_z^2)\sigma \\
                     & =-\sum_{q=0}^2A_q(\hat{S}_z)^q
    \end{aligned}
    \label{ParaHam}
\end{align}
where $A_0\equiv \lambda\sigma$, $A_1\equiv g\muB\mu_0H$ and $A_2\equiv K-\lambda\sigma$.
The partition function in the Spin-Coherent States basis $\ket{z}$ is given by
\begin{align}
    {\cal Z}&=\int d\mu(z)\bra{z}e^{-\beta \hat{\cal H}}\ket{z},
    \label{ParaHamPart}
\end{align}
where the measure $d\mu(z)$ is introduced to ensure the resolution of unity and $\beta=\frac{1}{k_B T}$. Here $k_B$ is Boltzmann's constant and $T$ is the temperature in Kelvin and $z\in \mathbb{C}$.
A spin coherent state $\ket{z}$ can be introduced for any principal quantum number $s$, from a reference state $\ket{0}$, as
\begin{align}
    \left\{
    \begin{array}{lcl}
        \ket{p}\!\! & \equiv & \!\!\ket{s,m_s=s-p} \\
        \ket{0}\!\! & \equiv & \!\!\ket{s,m_s=s}
    \end{array}
    \right.
\end{align}
where $\ket{s,m_s}$ is the {\it standard} spin basis, using the ladder operator $\hat{S}_-$ that lowers the eigenstate $m_s$, such that
\begin{align}
    \ket{z}\equiv(1+|z|^2)^{-s}\exp\left(z\hat{S}_-\right)\ket{0} & =(1+|z|^2)^{-s}\sum_{p=0}^{2s}
    \begin{pmatrix}
        2s \\
        p
    \end{pmatrix}^{\frac{1}{2}}z^p\ket{p},
    \label{SCS}
\end{align}
It is noteworthy that one can define spin coherent states in a similar fashion with either raising, lowering operators or a linear combination of both~\cite{radcliffePropertiesCoherentSpin1971}.

The action of the spin operators $\hat{S}_z$ and $\hat{\bm{S}}^2$ on the eigenstates $\ket{p}$ are given by
\begin{align}
\left\{
    \begin{matrix}
        \hat{S}_z\ket{p}=(s-p)\ket{p}\\
        \hat{S}^2\ket{p}=s(s+1)\ket{p}
    \end{matrix}
    \right.
\end{align}

To rewrite the partition function, it is useful to remember that the spin states constitute a complete set of basis vectors, in which the identity operator can be resolved as follows:
\begin{align}
    \mathds{1}&=\sum_{p=0}^{2s}\ket{p}\bra{p}
\end{align}
Upon introducing this expression in eq.\eqref{ParaHamPart}  one gets
\begin{align}
    {\cal Z}&=\int d\mu(z)\sum_{p=0}^{2s}\bra{z}e^{-\beta \hat{\cal H}}\ket{p}\braket{p|z}.
\end{align}
Taking advantage of the fact that the $\ket{p}$ states are eigenstates of the Hamiltonian given by eq.\eqref{ParaHam} one deduces that
\begin{align}
        {\cal Z}&=\int d\mu(z)\sum_{p=0}^{2s}|\braket{z|p}|^2\exp\left(\beta\sum_{q=0}^2A_q(s-p)^q\right).
\end{align}
Now, rewriting the expression for the Spin-Coherent State given by eq.\eqref{SCS} 
\begin{align}
        {\cal Z}&=\int \frac{d\mu(z)}{(1+|z|^2)^{2s}}\sum_{p=0}^{2s}\left(|z|^2\right)^p\begin{pmatrix}
            2s\\ p
        \end{pmatrix}\exp\left(\beta \sum_{q=0}^2A_q(s-p)^q\right),
\end{align}
 one can factorize the partition function as follows: 
\begin{align}
    {\cal Z}&=\exp\left(\beta\sum_{q=0}^2A_qs^q\right) \int \frac{d\mu(z)}{(1+|z|^2)^{2s}}L(2s, \beta, |z|^2)
    \label{partitionAlmostOk}
\end{align}
where $L(2s, \beta, |z|^2)$ is a polynomial in $|z|^2:$
\begin{align}
    L(2s, \beta, |z|^2)\equiv\sum_{p=0}^{2s}
    \begin{pmatrix}
        2s\\ p
    \end{pmatrix}
    \left(|z|^{2}\right)^p\exp\left(\beta A_2 p^2\right)\exp\left(-\beta\left(2sA_2+A_1\right)p\right).\label{notClosedForm}
\end{align}

A closed expression for $L(2s, \beta, |z|^2),$ for any value of $s$ and $z,$ would, certainly, be desirable. However, the term $\exp\left(\beta A_2p^2\right)$ complicates matters and prevents the use of the binomial formula
\begin{align}
    \sum_{p=0}^{2s}
    \begin{pmatrix}
            2s\\ p
    \end{pmatrix}1^{2s-p}\left(|z|^2\exp\left(\beta Q(s)\right)\right)^p&=(1+|z|^2\exp\left(\beta Q(s)\right))^{2s},
\end{align}
for values of any function $Q(s)$.

Therefore, there is no clear method to compute such a closed form explicitly.
It is possible, however, to compute the expression $L(2s, \beta, |z|^2)$ for  any value of $s$ and selected expressions are given in appendix~\ref{app:expressions}.

It is noteworthy that, when all the mechanical energy along the $z$ axis is opposite to the energy along the easy axis, i.e. $K=\lambda\sigma$, then $A_2=0$.
In this case, $L(2s,|z|^2)$ can be expressed in closed form-this represents the behaviour of a pure Zeeman paramagnet, and this indeed, can be mapped to the case studied in ref.~\cite{nussleNumericalSimulationsSpin2023}.

Following the path of this previous work of ours, our aim is now to obtain an effective Hamiltonian for performing classical simulations. To this end,  one needs to express the integrand of eq.~\eqref{partitionAlmostOk} in exponential form.

We start by writing 
\begin{equation}
    \frac{L(2s,|z|^2)}{\left(1+|z|^2\right)^{2s}}\equiv\exp\left(-\beta {\cal H}_{\textrm{eff}}(\beta,2s,|z|^2)\right).
\end{equation}
This expression defines the effective Hamiltonian, as it yields
\begin{equation}
    {\cal H}_{\textrm{eff}}(\beta,2s,|z|^2)=\frac{-1}{\beta}\ln\left(\frac{L(2s,|z|^2)}{\left(1+|z|^2\right)^{2s}}\right),
\end{equation}

In order to obtain the dominant terms that govern the behaviour of the system at ``high temperatures", one can perform a Taylor expansion of the effective Hamiltonian ${\cal H}_{\textrm{eff}}(\beta,2s,|z|^2)$ in the limit $\beta\rightarrow 0$.

\begin{equation}
    {\cal H}^{\textrm{high-T}}_\textrm{eff}(\beta,2s, |z|^2,N)\equiv\sum_{j=0}^N\frac{\beta^j}{j!}\left.\frac{\partial{\cal H}_{\textrm{eff}}(\beta,2s,|z|^2)}{\partial \beta^j}\right|_{\beta=0}\label{taylorS}
\end{equation}
where $N$ is the order of the Taylor expansion.
This allows us to derive polynomial expressions in terms of increasing powers of $\beta$, that provide insight into the properties of the paramagnetic system at temperatures greater than $\langle {\cal H}_{\textrm{eff}}(0,2s,|z|^2)/k_b\rangle$. 
In fact, this expansion remains meaningful, as long as ${\cal H}_{\textrm{eff}}(0,2s,|z|^2)\beta\ll 1$. 
For example, if the dominant term of \eqref{ParaHam} is the anisotropy, then $Ks\beta\ll 1$ must be satisfied.
For convenience, the terms in expression \eqref{taylorS} are relabeled as
\begin{equation}
    {\cal H}^{\textrm{high-T}}_\textrm{eff}(\beta,2s, |z|^2,N)\equiv\sum_{j=0}^N\beta^j{\cal H}^{(j)}_\textrm{eff}(2s, |z|^2)\label{effHamiltonianAlgo},
\end{equation}
hence we have
\begin{align}
    {\cal H}^{(j)}_\textrm{eff}(2s, |z|^2)&=\frac{1}{j!}\left.\frac{\partial{\cal H}_{\textrm{eff}}(\beta,2s,|z|^2)}{\partial \beta^j}\right|_{\beta=0}
\end{align}
and expressions for ${\cal H}^{(j)}_\textrm{eff}(2s, |z|^2)$ are derived explicitly for several values of $s$ and for small $j$ in appendix~\ref{app:expressions}.
The numerical compendium \cite{thisDataset} includes a Python software package that can be used for calculating the effective Hamiltonian \eqref{effHamiltonianAlgo} for any given values of $s$ and $N$.

Now, using the mapping, $z:\mathbb{C}\to S^2,$
from the complex plane $z\in\mathbb{C}$ to the unit sphere, $\bf u\in S^2$~\cite{karchevPathIntegralRepresentation2012}; the component $u_z$ of the unit vector is related to $|z|$ by 
\begin{equation}
    |z|^2=\frac{1-u_z}{1+u_z}.\label{stereog}
\end{equation}
To perform spin dynamics,  one needs to calculate the effective field, which is simply the functional derivative of the effective Hamiltonian (for the high temperature Taylor series approach, the relevant effective Hamiltonian will be ${\cal H}^{\textrm{high-T}}_\textrm{eff}$) \cite{erikssonAtomisticSpinDynamics2017} along a particular direction $\bm u$ on the 2-sphere
\begin{equation}
    -\frac{1}{g\muB s}\frac{\partial {\cal H}_\textrm{eff}}{\partial \bm u}=\bm{B}_\textrm{eff}
    \label{B_eff}.
\end{equation}
Here, this leads to an effective field along the $z$-direction only.

Because of eq.~\eqref{effHamiltonianAlgo}, the effective field can, also, be usefully expanded in powers of the inverse of the temperature as
\begin{align}
    B^z_{\textrm{eff}}(\beta,u_z, N)&=\sum_{j=0}^N\beta^{j} B_{\textrm{eff}}^{(j)}(u_z)
    \label{B_eff_expansion}
\end{align}
and we provide sample expressions in appendix~\ref{app:expressions}.  The provided numerical compendium also enables the computation of this effective field.

Having computed the effective field, we will now introduce it in a standard atomistic spin dynamics simulation, but first, some analytical expressions to compare our results to are required. Thus we will now derive exact expressions from the partition function in the quantum case.

\section{Exact quantum cumulants}
\label{sec:quantumResults}

For a single spin in thermal equilibrium, where the quantum Hamiltonian given by eq.~\eqref{ParaHam} is a function of $\hat{S}_z$ only, one can compute cumulants directly from the partition function in the standard spin states basis $\ket{s,m}$, which is given by
\begin{equation}
    \begin{aligned}
        {\cal Z} &=\sum_{m_s=-s}^{s}\bra{s,m_s}e^{-\beta \hat{\cal H}}\ket{s,m_s}\\
        &=\sum_{m_s=-s}^{s}\exp\left(\beta \left(A_0 + A_1 m_s + A_2 m_s^2\right)\right)\label{spinStatePartition}
    \end{aligned}
\end{equation}
Consequently the expression for the first-order cumulant for the $z$-component of the spin operator, which is equivalent to the first order moment $\braket{\hat{S}_z}$, that can be compared directly to the projection along this axis of the magnetization of non-interacting spins, is computed as
\begin{align}
	\begin{aligned}
        \braket{\braket{\hat{S}_z}}&=\frac{1}{\cal Z}\sum_{m_s=-s}^{s}\bra{s,m_s}\hat{S}_ze^{-\beta \hat{\cal H}}\ket{s,m_s}\\
	&=\frac{1}{\cal Z}\sum_{m_s=-s}^{s}m_s\exp\left(\beta \left(A_0 + A_1 m_s + A_2 m_s^2\right)\right).
\end{aligned}\label{analyticalExpectationofSz}
\end{align}
Fluctuations about this average value (in our case both thermal and quantum) can be evaluated by computing the higher order moments; the first example thereof is the second-order cumulant
\begin{align}
    \begin{aligned}
        \braket{\braket{\hat{S}^2_z}}&=\braket{\hat{S}_z^2}-\braket{\braket{\hat{S}_z}}^2\\
        &=\frac{1}{\cal Z}\sum_{m_s=-s}^{s}m_s^2\exp\left(\beta \left(A_0 + A_1 m_s + A_2 m_s^2\right)\right)-\braket{\braket{\hat{S}_z}}^2
    \end{aligned}
\end{align}
When $\braket{\hat{S}_z}$ does not vanish, the relevant quantity that is useful for probing the thermal scaling of these fluctuations \cite{schonhammerQuantumThermalFluctuations2014} is then given by $\frac{\sqrt{\braket{\braket{\hat{S}^2_z}}}}{\braket{\hat{S}_z}}$.
This represents a benchmark for the domain of temperatures where quantum fluctuations become relevant, for any ``quantum-corrected" Hamiltonian.
This will be the subject of future work.

These expressions can be compared to successive approximations in the high-temperature limit, by providing the corresponding effective field computed by eq.~\eqref{B_eff_expansion} to classical atomistic spin dynamics simulations, once the system reaches equilibrium.
This is the subject of the next section. 

\section{Results from atomistic simulations}
\label{sec:results}

When performing atomistic spin dynamics simulations, the main object of interest is the real-time dynamics of a magnetic moment $\bm{m}$ in a given effective magnetic field $\bm{B}_\textrm{eff}$. 
The evolution of such a dynamics is given by a dampened precession equation, a.k.a. the Landau-\-Lifshitz-\-Gilbert equation~\cite{erikssonAtomisticSpinDynamics2017}
\begin{equation}
	\dot{\bm{m}}=-\frac{\gamma}{1+\alpha^2}\left(\bm{m}\times\bm{B}_\textrm{eff}+\alpha\bm{m}\times\left(\bm{m}\times\bm{B}_\textrm{eff}\right)\right)\label{LLG}
\end{equation}
where $\gamma=\frac{g\muB}{\hbar}$ is the gyromagnetic ratio in $\textrm{rad}\,\textrm{s}^{-1}\,\textrm{T}^{-1}$ and $\alpha$ is the dimensionless Gilbert damping parameter. The effective field is assumed to be defined from an effective Hamiltonian ${\cal H}_\textrm{eff}$, which encodes the interactions of the system at hand
\begin{equation}
	\bm{B}_\textrm{eff}=-\frac{1}{\mu_s}\frac{\partial {\cal H}_\textrm{eff}}{\partial \bm{m}}
\end{equation}
where $\mu_s=g\muB s$ is the length of the spin magnetic moment. 

To include temperature in this formalism, one supplements the effective magnetic field by a stochastic noise $\bm{\eta}$
\begin{equation}
	\bm{B}_\textrm{eff}\rightarrow\bm{B}_\textrm{eff}+\bm{\eta}
\end{equation}
such as equation \eqref{LLG} becomes a stochastic differential equation with a multiplicative noise that satisfies the Fluc-\-tuation-\-Dissipation theorem~\cite{bertottiNonlinearMagnetizationDynamics2009}. The stochastic noise is defined by its moments which in the classical case define a white noise
\begin{equation}
    \label{noise_props}
\left\{
	\begin{matrix}
		\braket{\eta_i(t)}&=&0\\
		\braket{\eta_i(t)\eta_j(t')}&=&\displaystyle\frac{2\alpha}{\beta\mu_s\gamma}\delta_{ij}\delta(t-t')
	\end{matrix}
	\right.
\end{equation}
where $i$ and $j$ are cartesian coordinates components. All the cumulants higher and equal to 3 are taken to be zero.
It is noteworthy that, at this point, in eq.~\eqref{noise_props} any colored-noise expression can be considered to fully describe the physical nature of the (external) bath. 
What we want to do is identify the average values of the magnetization, computed by solving the Landau-Lifshitz-Gilbert equation, in presence of the noise field ${\bm \eta},$
with  the quantum cumulants computed in section~\eqref{sec:quantumResults}. 
To this end,  one must check for ergodicity, upon integrating equation \eqref{LLG} numerically using a symplectic integration scheme \cite{thibaudeauThermostattingAtomicSpin2012} for $N_s$ realisations (hence independent trajectories) of the noise.
In this approach, therefore, the average magnetization, which is to be compared to eq.~\eqref{analyticalExpectationofSz} is given by the expression
\begin{equation}
	\braket{m_z}=\frac{1}{N_s}\frac{1}{N_t}\sum_{i=1}^{N_s}\sum_{j=1}^{N_t}m_z^{(i)}(t_j)
\end{equation}
with $N_t$ is the number of time samples. 
The computation of any expectation values are performed after an initial equilibration period of $5$ ns (in the units used) in order to let the system relax to a well defined thermalised state, as the system requires a few nanoseconds to equilibrate. 
This can be monitored by computing the instantaneous microcanonical spin-temperature~\cite{nurdinDynamicalTemperatureSpin2000} to ensure that convergence to equilibrium is achieved. 

\begin{figure}[htbp]
    \centering
    \begin{tabular}{cc}
        \includegraphics[width=0.5\textwidth]{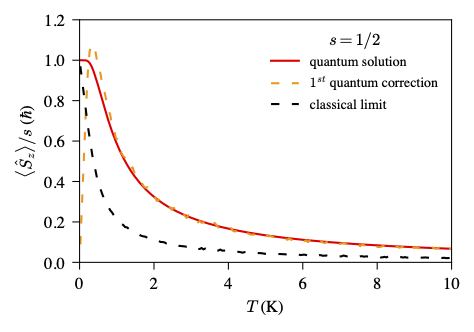}&
        \includegraphics[width=0.5\textwidth]{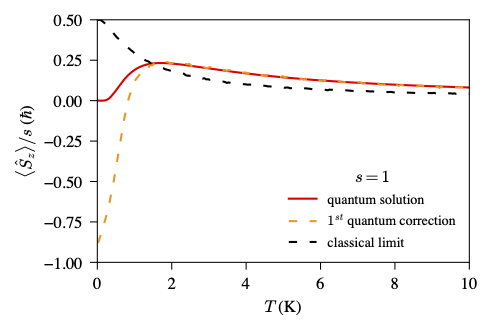}\\
        \includegraphics[width=0.5\textwidth]{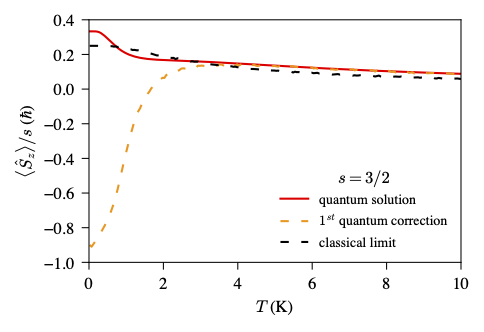}&
        \includegraphics[width=0.5\textwidth]{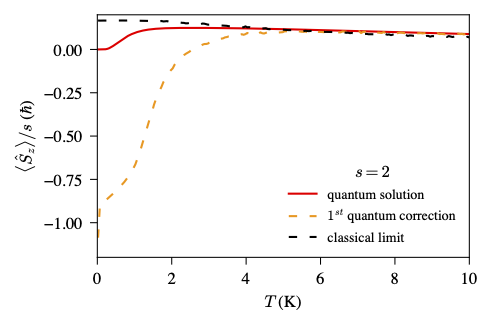}
    \end{tabular}
    \caption{Expectation value for $z$-component of spin as a function of temperature with $\mu_0H_z=1T$, $K=-2g\muB\mu_0 H_z$ and $\lambda\sigma=0$ for $s=\left\{1/2,1,3/2,2\right\}$. Analytical quantum result using \eqref{analyticalExpectationofSz} (red solid-line), classical limit (black dashed-line) and first quantum correction using quantum ASD (orange dashed-line)}
    \label{fig:ZeemanParamagnet}
\end{figure}

The time we take for the average itself is as long as $15$ ns for a constant integration timestep of $5.10^{-5}$ ns. We repeat this procedure over $N_s=20$ realisations of the noise. The initial condition for the orientation of the magnetic moment in the atomistic simulations is
\begin{equation}
    \bm{m}_0=\frac{1}{\sqrt{3}}\left(\begin{matrix}
         & 1 \\
         & 1 \\
         & -1
    \end{matrix}\right)
\end{equation}
which indeed satisfies $||\bm{m}_0||=1$.
We use the effective fields derived in section \ref{sec:definition} in order to compute approximate thermal expectation of the quantum system described by Hamiltonian \eqref{ParaHam} from this effective classical method. Here we identify the unit vector $\bm{u}$ which appeared in the stereographic projection mapping \eqref{stereog} from our derivation, with the unit vector for the magnetisation, $\bm{m}$. Results for several values of $s$ are given in figure \ref{fig:ZeemanParamagnet}. In this case, the effective Hamitonian is given by \eqref{effHamiltonianAlgo}. The classical limit is equivalent to taking $N=0$, the first quantum correction to $N=1$, and the exact quantum solution is computed analytically using \eqref{spinStatePartition} and \eqref{analyticalExpectationofSz}.

One can see here that the threshold temperature for accurate simulation increases with increasing value of $s$, which agrees with the condition where, looking at the effective field provided in \eqref{eff_Hamilton_dominant} one can see that see that the relevant scaling is $\frac{s(2s-1)}{2}\frac{K}{k_B}\ll T\Leftrightarrow T\gg s(2s-1)1.34 K$. For spin $s=1/2$ and $s=1$  it is remarkable that, not only does the effective model correctly describe even very low temperatures of the thermal expectation values, but it also captures the different features of half integer vs integer spin behaviour with the initial maximal value for half integer spins, and minimal value for integer spins. However, for higher spins, as the low temperature range is not correctly approximated, these features are lost.

\begin{figure}
    \centering
    \begin{tabular}{cc}
        \includegraphics[width=0.5\textwidth]{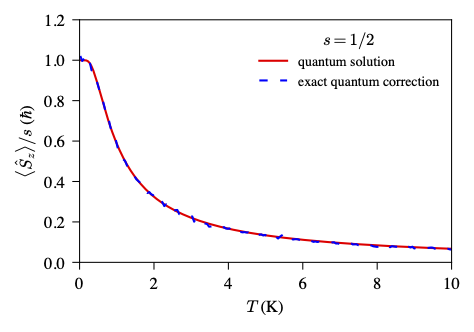}&
        \includegraphics[width=0.5\textwidth]{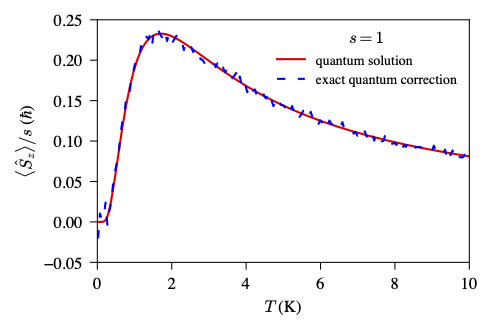}\\
        \includegraphics[width=0.5\textwidth]{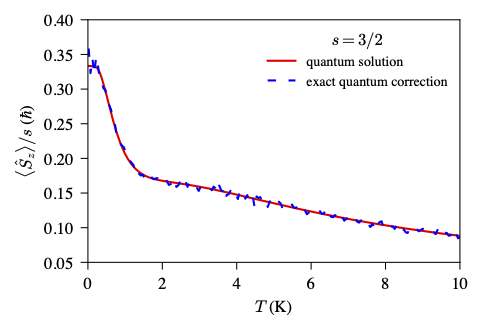}&
        \includegraphics[width=0.5\textwidth]{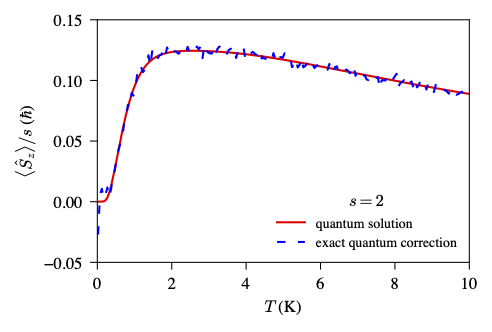}
    \end{tabular}
    \caption{Expectation value for $z$-component of spin as a function of temperature with $\mu_0H_z=1T$, $K=-2g\muB\mu_0 H_z$ and $\lambda\sigma=0$ for $s=\left\{1/2,1,3/2,2\right\}$. Analytical quantum result using \eqref{analyticalExpectationofSz} (red solid-line), quantum correction using quantum ASD (blue dashed-line) with the {\it exact} Hamiltonian \eqref{exactHam}}
    \label{fig:quantum-exact-ASD}
\end{figure}

To implement a model that is valid for the whole temperature range, an obvious idea is to skip the high-temperature Taylor expansion from \eqref{effHamiltonianAlgo} and simply take as our effective Hamiltonian, the following expression
\begin{equation}
	{\cal H}_\textrm{eff}(\beta, 2s, |z|^2)\equiv -\frac{1}{\beta}\ln\left(\frac{L(2s,|z^2|)}{\left(1+|z|^2\right)^{2s}}\right)\label{exactHam}
\end{equation}
Indeed, technically, nothing prevents us from taking the derivative of this Hamiltonian in order to compute an effective field for our atomistic simulations (a selection of detailed expressions are given in appendix~\ref{app:exactLogHam}). In doing so, we obtain results valid over the whole temperature range as seen in figure \ref{fig:quantum-exact-ASD} . These are the results labeled as {\it exact quantum correction} on figure \ref{fig:quantum-exact-ASD}. The price paid for this, however, is that the Hamiltonian and the effective field, are completely different expressions from what the classical limit would be, and often contain numerically more expensive functions, instead of simple polynomials. Moreover, this effective field contains a denominator that is potentially singular, as a function of the spin components. In practice, however, these potential singularities are never reached in the numerical methods and hence are straightforward to deal with.

\begin{figure}
    \centering
    \begin{tabular}{cc}
        \includegraphics[width=0.5\textwidth]{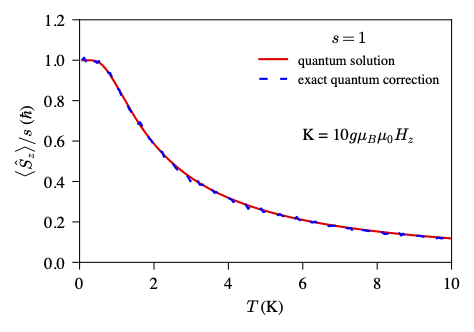}&
        \includegraphics[width=0.5\textwidth]{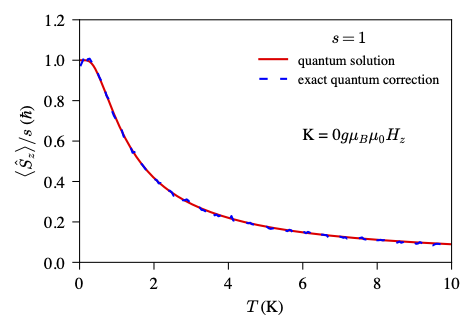}\\
        \includegraphics[width=0.5\textwidth]{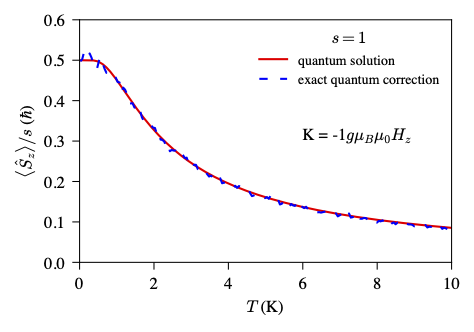}&
        \includegraphics[width=0.5\textwidth]{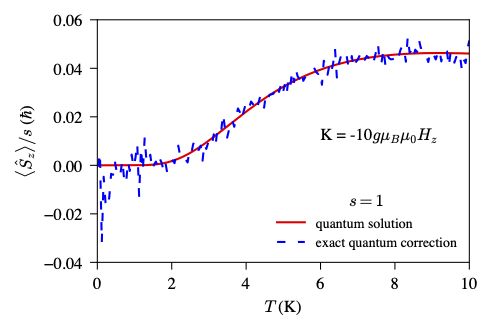}
    \end{tabular}
    \caption{Expectation value for $z$-component of spin as a function of temperature with $\mu_0H_z=1T$, $s=1$ and $\lambda\sigma=0$ for $K=\{10,0,-1,-10\}g\muB\mu_0 H_z$ . Analytical quantum result using \eqref{analyticalExpectationofSz} (red solid-line), quantum correction using quantum ASD (blue dashed-line) with the {\it exact} Hamiltonian \eqref{exactHam}}
    \label{fig:benchmark}
\end{figure}

It is interesting to notice that this method is indeed able to correctly reproduce the different characteristic behaviour as a function of the strength and sign of the anisotropy constant $K$ (positive means easy axis, negative hard axis) as shown in figure \ref{fig:benchmark}. Indeed at first, the easy axis ($K = 10 g\mu_B\mu_0H_z$) simply makes the alignment along the field easier and thermally more stable than only with the Zeeman term ($K=0$). Once the sign changes, the maximum alignment with the field is reduced ($K = -1 g\mu_B\mu_0H_z$) before it finally becomes energetically more favourable to be perpendicular to the hard axis at low temperatures ($K = -2 g\mu_B\mu_0H_z$ in figure \ref{fig:quantum-exact-ASD}). after this point, increasing the strength of the anisotropy ($K = -10 g\mu_B\mu_0H_z$) compresses the overall expectation value closer to 0 and moves the minimal thermal fluctuations to overcome the anisotropy higher. This behaviour is well known from an exact quantum mechanical perspective, but is here reproduced by means of a classical simulation in terms of a magnetic moment, which is our ultimate goal.

\section{Conclusions}
\label{sec:conclusions}

In this work we have presented a method for computing quantum thermal expectation values of spin systems using an effective classical atomistic spin dynamics simulation where the effective field is computed from the quantum partition function in the spin coherent states basis. 
This work generalises the approach of the paper \cite{nussleNumericalSimulationsSpin2023} by supplementing the Zeeman Hamiltonian by an external strain and an uniaxial anisotropy along the quantisation $z-$axis. Using this method, we have been able to produce thermal expectation values following two slightly different approximation schemes, namely a high temperature approach where the Hamiltonian is expanded in powers or $\beta$, and a method which computes the effective atomistic magnetic field directly from the Hamiltonian, hence providing an accurate description over the whole range of temperatures. 

The quantum system is exactly solvable, and serves as a proof of concept for a method that can also be implemented in much larger --potentially interacting-- spin systems, as will be the focus of future work. This allows a multiscale approach in the sense that it relies upon first deriving the effective field within a fully quantum mechanical framework and then introducing this effective field in a large-scale atomistic simulation, where the fluctuations about the average value are described through the effects of the noise fields. 

In the high temperature approximation, we have demonstrated that the essential characteristics of thermal expectation values can be accurately represented even for relatively low $s$ values. This is due to the dominant term of the Hamiltonian scaling with the principal quantum number in our given example. Furthermore, we have provided evidence  that even without a straightforward polynomial expression for the Hamiltonian, it is feasible to calculate an effective field for our atomistic simulations. This allows us to probe  the entire temperature range and even reproduce the  significant differences between half-integer and integer spin values. What is remarkable is that  all this can be  achieved while maintaining an effective classical approach. 
\appendix

\section{Explicit expressions for selected values, high-temperature}
\label{app:expressions}
In this section, we compute the values of the expressions $L(2s,|z|^2)$, $H_\textrm{eff}(2s,u_z)$ and $B_\textrm{eff}^z(\beta,2s,u_z)$ given in section~\ref{sec:definition}, for $s\in[\frac{1}{2},1,\frac{3}{2},2]$

By increasing $s$ by $1/2$ increments, the first three values of $L$ are
\begin{align}
    \begin{aligned}
        L(1,|z|^{2})&=|z|^{2} e^{- A_{1} \beta} + 1\\
        L(2,|z|^{2})&=|z|^{4} e^{- 2 A_{1} \beta} + 2 |z|^{2} e^{- A_{1} \beta - A_{2} \beta} + 1\\
        L(3,|z|^{2})&=|z|^{6} e^{- 3 A_{1} \beta} + 3 |z|^{4} e^{- 2 A_{1} \beta - 2 A_{2} \beta} + 3 |z|^{2} e^{- A_{1} \beta - 2 A_{2} \beta} + 1\\
        L(4,|z|^{2})&=|z|^{8} e^{- 4 A_{1} \beta} + 4 |z|^{6} e^{- 3 A_{1} \beta - 3 A_{2} \beta} + 6 |z|^{4} e^{- 2 A_{1} \beta - 4 A_{2} \beta} + 4 |z|^{2} e^{- A_{1} \beta - 3 A_{2} \beta} + 1
    \end{aligned}
\end{align}

The leading terms of the effective Hamiltonian, considered not as a function of the temperature, are given by
\begin{align}
    \begin{aligned}
        s=\frac{1}{2}\hspace{5mm}{\cal H}^{(0)}_\textrm{eff}(|z|^2)&=\frac{A_{1} |z|^{2}}{|z|^{2} + 1}\\
        s=1\hspace{5mm}{\cal H}^{(0)}_\textrm{eff}(|z|^2)&=\frac{2 |z|^{2} \left(A_{1} |z|^{2} + A_{1} + A_{2}\right)}{|z|^{4} + 2 |z|^{2} + 1}\\
        s=\frac{3}{2}\hspace{5mm}{\cal H}^{(0)}_\textrm{eff}(|z|^2)&=\frac{3 |z|^{2} \left(A_{1} |z|^{2} + A_{1} + 2 A_{2}\right)}{|z|^{4} + 2 |z|^{2} + 1}\\
        s=2\hspace{5mm}{\cal H}^{(0)}_\textrm{eff}(|z|^2)&=\frac{4 |z|^{2} \left(A_{1} |z|^{2} + A_{1} + 3 A_{2}\right)}{|z|^{4} + 2 |z|^{2} + 1}
    \end{aligned}
    \label{effHam}
\end{align}

Here are the expressions of eq.~\eqref{effHam} as functions of $u_z$,
\begin{align}
	    \forall s \hspace{5mm} {\cal H}^{(0)}_\textrm{eff}(u_z)&=A_1 s (1-u_z)+A_2\frac{s(2s-1)}{2}(1-u_z^2)\label{eff_Hamilton_dominant}
\end{align}

Here are the leading terms of eq.~\eqref{B_eff} as functions of $u_z$,
\begin{align}
    \forall s \hspace{5mm} B^{(0)}_\textrm{eff}(u_z)&=\frac{A_1+(2s-1)A_2u_z}{g\mu_B} \label{eff_field_dominant}
\end{align}

These terms all correspond to the classical limit of the Hamiltonian \eqref{ParaHam}. If  one wants to take quantum corrections, one requires to go to the first order of non-vanishing $\beta$. We shall take the example of spin 1 (as $A_2$ contributions vanish for spin $1/2$), but any spin to any approximation order (given enough time and computational power), can be computed using the python package.
\begin{equation}
	s=1\hspace{5mm}{\cal H}^{(1)}_\textrm{eff}(|z|^2)=\frac{ |z|^{2} \left(- A_{1}^{2} |z|^{4} - 2 A_{1}^{2} |z|^{2} - A_{1}^{2} + 2 A_{1} A_{2} |z|^{4} - 2 A_{1} A_{2} - A_{2}^{2} |z|^{4} - A_{2}^{2}\right)}{|z|^{8} + 4 |z|^{6} + 6 |z|^{4} + 4 |z|^{2} + 1}
\end{equation}
This expression can again be remapped onto the unit sphere
\begin{equation}
	s=1\hspace{5mm}{\cal H}^{(1)}_\textrm{eff}(u_z)=\frac{ \left(2 A_{1}^{2} u_{z}^{2} - 2 A_{1}^{2} + 4 A_{1} A_{2} u_{z}^{3} - 4 A_{1} A_{2} u_{z} + A_{2}^{2} u_{z}^{4} - A_{2}^{2}\right)}{8}
\end{equation}
Finally yielding the following expression for the effective field
\begin{equation}
	s=1\hspace{5mm}B^{(1)}_\textrm{eff}(u_z) =\frac{ \left(- A_{1}^{2} u_{z} - 3 A_{1} A_{2} u_{z}^{2} + A_{1} A_{2} - A_{2}^{2} u_{z}^{3}\right)}{2 g\mu_{B}}
\end{equation}

To build the corresponding effective field (recalling that only the $z$-component is non vanishing in our case) for the atomistic simulations, one simply needs to identify $\bm{u}$ with the magnetic moment $\bm{m}$ and build 
\begin{equation}
    \bm{B}_\textrm{eff}(\beta, u_z, N)=\left(\sum_{j=0}^N\beta^jB_\textrm{eff}^{(j)}(u_z)\right)\bm{e}_z
\end{equation}
where $\bm{e}_z$ is a unit vector along the $z$-direction.

\section{Explicit expressions for the s=1 exact Hamiltonian}
\label{app:exactLogHam}

Instead of performing a high temperature Taylor expansion to obtain the Hamiltonian, if one takes the expression \eqref{exactHam} then
\begin{align}
    \begin{aligned}
        s=1\hspace{5mm}\tilde{{\cal H}}_\textrm{eff}(|z|^2, \beta)&=2 A_{1} + A_{2} + \frac{1}{\beta}\log{\left(\frac{\left(|z|^{2} + 1\right)^{2}}{|z|^{4} e^{A_{2} \beta} + 2 |z|^{2} e^{A_{1} \beta} + e^{\beta \left(2 A_{1} + A_{2}\right)}} \right)}\\
        \tilde{{\cal H}}_\textrm{eff}(u_z, \beta)&=2 A_{1} + A_{2} + \frac{1}{\beta}\log{\left(\frac{4}{\left(u_{z} - 1\right)^{2} e^{A_{2} \beta} - 2 \left(u_{z} - 1\right) \left(u_{z} + 1\right) e^{A_{1} \beta} + \left(u_{z} + 1\right)^{2} e^{\beta \left(2 A_{1} + A_{2}\right)}} \right)}
    \end{aligned}
\end{align}
once remapped onto the unit sphere and following the usual procedure \eqref{B_eff} one can deduce the corresponding effective field
\begin{equation}
	s=1\hspace{5mm}\tilde{\bm{B}}_\textrm{eff}(u_z, \beta)=\frac{2 \left(\left(1 - u_{z}\right) e^{A_{1} \beta} + \left(u_{z} - 1\right) e^{A_{2} \beta} - \left(u_{z} + 1\right) e^{A_{1} \beta} + \left(u_{z} + 1\right) e^{\beta \left(2 A_{1} + A_{2}\right)}\right)}{\mu_{B} \beta g \left(\left(u_{z} - 1\right)^{2} e^{A_{2} \beta} - 2 \left(u_{z} - 1\right) \left(u_{z} + 1\right) e^{A_{1} \beta} + \left(u_{z} + 1\right)^{2} e^{\beta \left(2 A_{1} + A_{2}\right)}\right)}\bm{e}_z
\end{equation}

\medskip

\section*{Data Access}
Python code and output data to reproduce all results and figures reported in this paper are openly available from the Zenodo repository: \textit{Sources for: Path integral spin dynamics for quantum paramagnets.} \url{https://doi.org/10.5281/zenodo.11072984}~\cite{thisDataset}. The repository contains:
\begin{itemize}
    \item Python code to generate analytic equations derived herein.
    \item Python code to perform enhanced atomistic spin dynamics calculations with the quantum effective fields.
    \item Python scripts to reproduce all figures.
\end{itemize}
The software and data are available under the terms of the MIT License.

\section*{Author Contributions}
T.N.: conceptualization, methodology, investigation, software, writing—original draft; P.T.: conceptualization, methodology, writing—review and editing; S.N.: methodology, writing—review and editing.

\section*{Acknowledgements} 
This work was supported by the Engineering and Physical Sciences Research Council [grant number EP/V037935/1]. T.N acknowledges funding from the Royal Society.

\bibliographystyle{MSP}
\bibliography{PISD.bib}

\begin{thebibliography}{10}
\providecommand{\url}[1]{\texttt{#1}}
\providecommand{\urlprefix}{URL }

\bibitem{laflorencieQuantumEntanglementCondensed2016}
N.~Laflorencie,
\newblock \emph{Physics Reports} \textbf{2016}, \emph{646} 1.

\bibitem{nelsonQuantumFluctuations1985}
E.~Nelson,
\newblock \emph{Quantum Fluctuations},
\newblock Princeton Series in Physics. Princeton University Press, Princeton,
  N.J, \textbf{1985}.

\bibitem{foulkesQuantumMonteCarlo2001}
W.~M.~C. Foulkes, L.~Mitas, R.~J. Needs, G.~Rajagopal,
\newblock \emph{Rev. Mod. Phys.} \textbf{2001}, \emph{73}, 1 33.

\bibitem{andersSpinPrecessionRealtime2006}
F.~B. Anders, A.~Schiller,
\newblock \emph{Phys. Rev. B} \textbf{2006}, \emph{74}, 24 245113.

\bibitem{erikssonAtomisticSpinDynamics2017}
O.~Eriksson, A.~Bergman, L.~Bergqvist, J.~Hellsvik,
\newblock \emph{Atomistic Spin Dynamics: Foundations and Applications},
\newblock Oxford university press, Oxford, \textbf{2017}.

\bibitem{tranchidaMassivelyParallelSymplectic2018}
J.~Tranchida, S.~J. Plimpton, P.~Thibaudeau, A.~P. Thompson,
\newblock \emph{J. Comp. Phys.} \textbf{2018}, \emph{372} 406.

\bibitem{barkerSemiquantumThermodynamicsComplex2019}
J.~Barker, G.~E.~W. Bauer,
\newblock \emph{Physical Review B} \textbf{2019}, \emph{100}, 14 140401(R).

\bibitem{berrittaAccountingQuantumEffects2023}
M.~Berritta, S.~Scali, F.~Cerisola, J.~Anders,
\newblock Accounting for {{Quantum Effects}} in {{Atomistic Spin Dynamics}},
  \textbf{2023}.

\bibitem{andersQuantumBrownianMotion2022}
J.~Anders, C.~R.~J. Sait, S.~A.~R. Horsley,
\newblock \emph{New Journal of Physics} \textbf{2022}, \emph{24}, 3 033020.

\bibitem{nussleNumericalSimulationsSpin2023}
T.~Nussle, S.~Nicolis, J.~Barker,
\newblock \emph{Physical Review Research} \textbf{2023}, \emph{5}, 4 043075.

\bibitem{dutremoletdelacheisserieMagnetostrictionTheoryApplications1993}
E.~{Du Tr{\'e}molet de Lacheisserie},
\newblock \emph{Magnetostriction: Theory and Applications of
  Magnetoelasticity},
\newblock CRC Press, Boca Raton, FL, \textbf{1993}.

\bibitem{radcliffePropertiesCoherentSpin1971}
J.~M. Radcliffe,
\newblock \emph{Journal of Physics A: General Physics} \textbf{1971}, \emph{4},
  3 313.

\bibitem{thisDataset}
T.~Nussle, S.~Nicolis, J.~Barker, P.~Thibaudeau,
\newblock {Sources for: Path integral spin dynamics for quantum paramagnets
  (v1.0.8) [Data set]}, \textbf{2024},
\newblock {Zenodo}.

\bibitem{karchevPathIntegralRepresentation2012}
N.~Karchev,
\newblock \emph{arXiv:1211.4509 [cond-mat]} \textbf{2012}.

\bibitem{schonhammerQuantumThermalFluctuations2014}
K.~Sch{\"o}nhammer,
\newblock \emph{American Journal of Physics} \textbf{2014}, \emph{82}, 9 887.

\bibitem{bertottiNonlinearMagnetizationDynamics2009}
G.~Bertotti, I.~D. Mayergoyz, C.~Serpico,
\newblock \emph{Nonlinear {{Magnetization Dynamics}} in {{Nanosystems}}},
\newblock Elsevier, \textbf{2009}.

\bibitem{thibaudeauThermostattingAtomicSpin2012}
P.~Thibaudeau, D.~Beaujouan,
\newblock \emph{Physica A: Statistical Mechanics and its Applications}
  \textbf{2012}, \emph{391}, 5 1963.

\bibitem{nurdinDynamicalTemperatureSpin2000}
W.~B. Nurdin, K.-D. Schotte,
\newblock \emph{Phys. Rev. E} \textbf{2000}, \emph{61}, 4 3579.

\end{thebibliography}
\end{document}